# Solution of the Dirac equation in a curved space with static metric

A. D. Alhaidari

*Saudi Center for Theoretical Physics, Jeddah 21438, Saudi Arabia*
*Physics Department, King Fahd University of Petroleum & Minerals, Dhahran 31261, Saudi Arabia*

**Abstract**: Compatibility of symmetric quantization of the Dirac equation in a curved space with general covariance gives a special representation of the spin connections in which their dot product with the Dirac gamma matrices becomes equal to the "covariant divergence" of the latter. Requiring that the square of the equation gives the conventional Klein-Gordon equation in a curved space results in an operator algebra for the Dirac gamma matrices that involves the "covariant derivative" connections and the Riemann-Christoffel connections. In 1+1 space-time with static metric, we obtain exact solutions of this Dirac equation model for some examples. We also formulate the interacting theory of the model with various coupling modes and solve it in the same space for a given potential configuration.

## I. INTRODUCTION AND FORMULATION

Recently [1], we have shown that unambiguous symmetric quantization of the Dirac wave operator, $\gamma^\mu p_\mu$, in $n+1$ dimensional curved space-time gives $i\hbar\gamma^\mu \partial_\mu + \frac{i}{2}\hbar(\partial_\mu \gamma^\mu)$, where $\{\gamma^\mu\}_{\mu=0}^n$ are the space-time dependent Dirac gamma matrices that are related to the metric tensor by $\{\gamma^\mu, \gamma^\nu\} = 2g^{\mu\nu}\mathbf{1}$ and repeated indices are summed over. Therefore, the conventional Fock-Weyl formulation of the Dirac equation in a curved space that reads $i\hbar\gamma^\mu(\partial_\mu + \Gamma_\mu)\Psi = mc\Psi$ [2] gets replaced by

$$\left[ i\hbar\gamma^\mu\left(\partial_\mu + \Gamma_\mu\right) + \tfrac{i}{2}\hbar\left(\partial_\mu\gamma^\mu\right)\right]\Psi = mc\Psi, \tag{1}$$

where $\{\Gamma_\mu\}$ are the spin connection matrices, $m$ is the rest mass of the spinor and $c$ is the speed of light. We can recover the standard representation by defining a new set of spin connections $\{\tilde{\Gamma}_\mu\}$ such that $\gamma^\mu\tilde{\Gamma}_\mu = \gamma^\mu\Gamma_\mu + \tfrac{1}{2}\partial_\mu\gamma^\mu$, which transforms Eq. (1) into

$$i\gamma^\mu\left(\partial_\mu + \tilde{\Gamma}_\mu\right)\Psi = m\Psi, \tag{2}$$

where we have adopted the relativistic units $\hbar = c = 1$. It is rather conceivable that the relation $\gamma^\mu\tilde{\Gamma}_\mu = \gamma^\mu\Gamma_\mu + \tfrac{1}{2}\partial_\mu\gamma^\mu$ could be satisfied with the hidden connections $\{\Gamma_\mu\}$ being a set of functions rather than matrices. In fact, for that we only need the projections of $\{\Gamma_\mu\}$ that are relevant to the dot product with the gamma matrices. If this is so and if we call these relevant projections $\{\Sigma_\mu\}$, then we can write $\gamma^\mu\tilde{\Gamma}_\mu =$ $\frac{1}{2} \partial_\mu \gamma^\mu + \Sigma_\mu \gamma^\mu$. Now, we propose our model of the Dirac equation in a curved space and parametrize it by replacing the factor $\frac{1}{2}$ multiplying the term $\partial_\mu \gamma^\mu$ with a dimensionless parameter $\lambda$. Consequently, we write the Dirac equation model as

$$i\left(\gamma^\mu \partial_\mu + \Omega\right)\Psi = m\Psi, \qquad (3)$$

where the space-time matrix $\Omega$ is equal to $\gamma^\mu \tilde{\Gamma}_\mu = \tilde{\nabla}_\mu \gamma^\mu$ and where the "covariant derivative" is defined as $\tilde{\nabla}_\mu = \lambda \partial_\mu + \Sigma_\mu$. For a complete specification of this one-parameter model we need to provide the set of space-time functions (projections) $\{\Sigma_\mu\}$. To do that, we start by requiring that the square of this Dirac equation gives the conventional Klein-Gordon equation,

$$g^{\mu\nu}\left(\partial_\mu \partial_\nu + \Gamma^\sigma_{\mu\nu} \partial_\sigma\right)\Psi = -m^2 \Psi, \qquad (4)$$

where $\Gamma^\mu_{\alpha\beta} = \frac{1}{2} g^{\mu\nu}\left(\partial_\alpha g_{\nu\beta} + \partial_\beta g_{\nu\alpha} - \partial_\nu g_{\alpha\beta}\right)$ are the Riemann-Christoffel connections and $\{g_{\mu\nu}\}$ are elements of the inverse of the metric tensor. One can show that this leads to the following matrix operator algebra for the Dirac gamma matrices that involves the connections $\{\Sigma_\mu\}$ and $\{\Gamma^\mu_{\alpha\beta}\}$

$$\slashed{\partial}\gamma^\mu = -\{\Omega, \gamma^\mu\} + G^\mu \mathbb{1}, \qquad (5a)$$

$$\slashed{\partial}\Omega = -\Omega^2, \qquad (5b)$$

where $\slashed{\partial} = \gamma^\mu \partial_\mu$ and $G^\mu = g^{\alpha\beta} \Gamma^\mu_{\alpha\beta}$. In the conventional Dirac equation, writing the spin connections as the most general decomposition in terms of the Dirac gamma matrices [3], one can show that the Riemann-Christoffel connections and the spin connections are related as follows [4]

$$\partial_\mu \gamma^\nu + [\Gamma_\mu, \gamma^\nu] + \Gamma^\nu_{\mu\sigma} \gamma^\sigma = 0. \qquad (6)$$

Compatibility of this relation with the matrix operator algebra (5a) gives a trivial representation of the spin connections as $\Gamma_\mu = g_{\mu\nu}\left(g^{\alpha\beta} \Gamma^\nu_{\alpha\beta}\right) = G_\mu$, which is a set of functions rather than matrices. This supports our speculation below Eq. (2). In the following section, we will assert this conclusion by example. Thus, the matrix nature of $\Omega$ is in the linear combination of the gamma matrices and their space-time divergence not in the spin connections. Now, the matrix operator algebra (5) together with the metric equation $\{\gamma^\mu, \gamma^\nu\} = 2g^{\mu\nu}\mathbb{1}$ constitute stringent requirements that pin down the representation of the Dirac gamma matrices and might avoid the non-uniqueness problem in writing the Dirac equation in a curved space [5].

It is worth noting that the Dirac equation model introduced here differs from the one proposed in [1] by two particulars. First, in the introduction of the projections (or connections) $\{\Sigma_\mu\}$ that were missing from the model in [1]. Second, the square of the equation in [1] does not give the conventional Klein-Gordon equation (4) but rather its canonical version where there is no first order derivatives. In the following section, we formulate our model in 1+1 space-time with static metric followed, in section III, by sample solutions of the model in the same space. In section IV, we introduce interaction by coupling the Dirac particle to various potentials and obtain an exact solution in this space for a given set of potentials.

## II. THE MODEL IN 1+1 SPACE-TIME WITH STATIC METRIC

In this space-time, the Dirac gamma matrices are 2×2 and functions only of the space coordinate $x$. The metric equation $\{\gamma^\mu, \gamma^\nu\} = 2g^{\mu\nu}\mathbf{1}$ dictates that the general form of the gamma matrices is $\begin{pmatrix} a & b \\ c & -a \end{pmatrix}$, where $\{a,b,c\}$ are independent functions of $x$. Additionally, $\Omega$ will also have the same form since $\Omega = (\lambda\partial_\mu + \Sigma_\mu)\gamma^\mu$. Consequently, the off-diagonal elements of the matrices on the right side of Eq. (5a) vanish. Using this fact on the left side of the equation for $\mu = 1$ leads to $\gamma^1 = a(x)\begin{pmatrix} 1 & \alpha \\ \beta & -1 \end{pmatrix}$, where $\alpha$ and $\beta$ are constants such that $\alpha\beta \neq -1$. For $\mu = 0$ we obtain $\gamma^0 = b(x)\begin{pmatrix} 1 & \alpha \\ \beta & -1 \end{pmatrix} + \begin{pmatrix} 0 & \rho \\ \tau & 0 \end{pmatrix}$, where $\rho$ and $\tau$ are two additional parameters. Thus, the associated space-time metric tensor has the following components

$$g^{11} = (1+\alpha\beta)a^2, \tag{7a}$$

$$g^{00} = (1+\alpha\beta)b^2 + (\alpha\tau + \beta\rho)b + \rho\tau, \tag{7b}$$

$$g^{01} = g^{10} = (1+\alpha\beta)ab + \tfrac{1}{2}(\alpha\tau + \beta\rho)a. \tag{7c}$$

The flat space limit, where the space-time metric is $g = \begin{pmatrix} 1 & 0 \\ 0 & -h^2 \end{pmatrix}$ and $h(x)$ is a real function, requires that

$$a(x)^2 \to -(1+\alpha\beta)^{-1}h(x)^2, \; b(x) \to -\tfrac{1}{2}\tfrac{\alpha\tau+\beta\rho}{1+\alpha\beta}, \text{ and } \rho\tau = 1 + \tfrac{1}{4}\tfrac{(\alpha\tau+\beta\rho)^2}{1+\alpha\beta}. \tag{8}$$

This suggests that $a(x)$ is pure imaginary and thus we replace it by $ia(x)$. Reality of the metric tensor element $g^{01}$ dictates that we should also replace $b(x)$ with $ib(x)$ provided that the parameter combination $\alpha\tau + \beta\rho$ is either zero or pure imaginary. For simplicity and without loss of generality, we choose $\alpha\tau = -\beta\rho$. Hence, the flat space limit dictates that $b(x) \to 0$ and $\rho\tau = 1$. Therefore, we obtain the following:

$$\gamma^1 = ia\begin{pmatrix} 1 & \alpha \\ \beta & -1 \end{pmatrix}, \; \gamma^0 = ib\begin{pmatrix} 1 & \alpha \\ \beta & -1 \end{pmatrix} + \begin{pmatrix} 0 & 1/\tau \\ \tau & 0 \end{pmatrix}, \tag{9}$$

$$g = \begin{pmatrix} 1 & 0 \\ 0 & 0 \end{pmatrix} - (1+\alpha\beta)\begin{pmatrix} b^2 & ab \\ ab & a^2 \end{pmatrix}, \tag{10}$$

where $\beta = -\alpha\tau^2$, $\alpha^2\tau^2 \neq 1$ and $\tau \neq 0$. Using this metric and its inverse, we can write $G_\mu = \tfrac{1}{2}g^{\alpha\beta}(\partial_\alpha g_{\mu\beta} + \partial_\beta g_{\mu\alpha} - \partial_\mu g_{\alpha\beta})$ and obtain the following

$$G_0 = (1+\alpha\beta)(ab' - a'b), \tag{11a}$$

$$G_1 = -\tfrac{a'}{a} + (1+\alpha\beta)\tfrac{b}{a}(a'b - ab') = -\tfrac{1}{a}(a' + bG_0), \tag{11b}$$

where the prime stands for the derivative with respect to $x$. Moreover, with $G^\mu = g^{\alpha\beta}\Gamma^\mu_{\alpha\beta} = g^{\mu\nu}G_\nu$, we obtain

$$G^0 = (1+\alpha\beta)ab', \; G^1 = (1+\alpha\beta)aa'. \tag{12}$$

In Eq. (5a), we substitute these $\{G^\mu\}$, the matrix $\Omega = i\lambda a'\begin{pmatrix} 1 & \alpha \\ \beta & -1 \end{pmatrix} + \Sigma_\mu\gamma^\mu$ and the Dirac gamma matrices given in (9). As a result, we obtain

$$\Sigma^0 = (1+\alpha\beta)(ab' + \lambda a'b), \; \Sigma^1 = (1+\lambda)(1+\alpha\beta)aa'. \tag{13}$$

Now, with $\Sigma_\mu = g_{\mu\nu}\Sigma^\nu$ and $\{g_{\mu\nu}\}$ being elements of the inverse of the metric tensor, we obtain

$$\Sigma_0 = (1+\alpha\beta)(ab' - a'b), \tag{14a}$$

$$\Sigma_1 = -(1+\lambda)\tfrac{a'}{a} + (1+\alpha\beta)\tfrac{b}{a}(a'b - ab') = -(1+\lambda)\tfrac{a'}{a} - \tfrac{b}{a}\Sigma_0. \tag{14b}$$

Using these findings, we can write the matrix $\Omega$ in the Dirac equation model (3) as

$$\Omega = -ia'\begin{pmatrix}1 & \alpha \\ \beta & -1\end{pmatrix} + \Sigma_0 \begin{pmatrix}0 & 1/\tau \\ \tau & 0\end{pmatrix}. \tag{15}$$

Therefore, we must conclude that in 1+1 space-time with static metric the model is $\lambda$ independent. However, this conclusion cannot be generalized to time-dependent metric or to other dimensional spaces. This is because in the matrix operator algebra (5) the term $\partial_\mu \gamma^\mu$ is not identically zero for all representations of the gamma matrices. Substituting from the results above in Eq. (5b) leads to the following two conditions:

(i) $d\Sigma_0/dx = 0$, meaning that $\Sigma_0$ is a constant of inverse length dimension, say $\eta$.

(ii) $a'^2 - aa'' = \eta^2/(1+\alpha\beta)$. (16)

The first condition implies that $a''/a = b''/b$, which means that both ratios are the same function, say $f(x)$. Hence, for a given $f(x)$ with an inverse square length dimension, the functions $a(x)$ and $b(x)$ are the two independent solutions of the following second order differential equation

$$y''(x) - f(x)y(x) = 0. \tag{17}$$

At this point, the formulation of the problem in 1+1 space-time with static metric is complete with the following steps:

(i) Given an appropriate function $f(x)$, obtain two independent solutions of Eq. (16) as $a(x)$ and $b(x)$. In writing these solutions, use the parameters $\{\alpha, \tau, \eta\}$ as much as possible else introduce additional parameters. Note that $\beta = -\alpha\tau^2$.

(ii) Use the equation $\Sigma_0 = (1+\alpha\beta)(ab' - a'b) = \eta$ and the constraint given by Eq. (16) together with the flat space limit to restrict the values of some of the problem parameters and/or establish a relationship among them.

With all ingredients of our Dirac equation model in 1+1 space-time with static metric identified, we can now formulate the general solution of the problem and then use it in the next section to present few examples. Writing the spinor wavefunction as $\Psi(t,x) = e^{-i\varepsilon t}\psi(x)$ and inserting the gamma matrices from Eq. (9) and the matrix $\Omega$ from Eq. (15) into the Dirac equation (3) gives the following matrix wave equation

$$\left\{\begin{pmatrix}1 & \alpha \\ \beta & -1\end{pmatrix}\left[i\varepsilon b(x) - a(x)\tfrac{d}{dx} + a'(x)\right] + (\varepsilon + i\eta)\begin{pmatrix}0 & 1/\tau \\ \tau & 0\end{pmatrix} - m\right\}\begin{pmatrix}\psi_\uparrow(x) \\ \psi_\downarrow(x)\end{pmatrix} = 0, \tag{18}$$

which could be rewritten as follows

$$\left\{-\begin{pmatrix}1 & \alpha \\ \beta & -1\end{pmatrix}a(x)\left[\tfrac{d}{dx} - i\varepsilon\tfrac{b(x)}{a(x)} - \tfrac{a'(x)}{a(x)}\right] + (\varepsilon + i\eta)\begin{pmatrix}0 & 1/\tau \\ \tau & 0\end{pmatrix} - m\right\}\begin{pmatrix}\psi_\uparrow(x) \\ \psi_\downarrow(x)\end{pmatrix} = 0. \tag{19}$$

Replacing $\psi(x)$ by $a(x)^{-1}\phi(x)$ in this equation will remove the last term inside the square bracket. If we also multiply from left by the constant matrix $\begin{pmatrix}1 & \alpha \\ \beta & -1\end{pmatrix}$ then we obtain

$$\left\{a(x)\sqrt{1+\alpha\beta}\left[\tfrac{d}{dx} - i\varepsilon\tfrac{b(x)}{a(x)}\right] + \tfrac{m-\alpha\tau E}{\sqrt{1+\alpha\beta}}\begin{pmatrix}1 & 0 \\ 0 & -1\end{pmatrix} + \tfrac{\alpha\tau m - E}{\sqrt{1+\alpha\beta}}\begin{pmatrix}0 & 1/\tau \\ -\tau & 0\end{pmatrix}\right\}\begin{pmatrix}\phi_\uparrow \\ \phi_\downarrow\end{pmatrix} = 0, \tag{20}$$

where $E = \varepsilon + i\eta$. Writing $\phi(x) = e^{i\varepsilon q(x)}\chi(x)$, where $\tfrac{dq}{dx} = b/a$, removes the second term inside the square bracket and transforms the equation into

–4–

$$\left[\frac{d}{dy}+\frac{m-\alpha\tau E}{\sqrt{1+\alpha\beta}}\begin{pmatrix}1 & 0\\ 0 & -1\end{pmatrix}+\frac{\alpha\tau m-E}{\sqrt{1+\alpha\beta}}\begin{pmatrix}0 & 1/\tau\\ -\tau & 0\end{pmatrix}\right]\begin{pmatrix}\chi_\uparrow\\ \chi_\downarrow\end{pmatrix}=0, \qquad (21)$$

where the new variable $y(x)$ solves the equation $\frac{dy}{dx}=1/a\sqrt{1+\alpha\beta}$. Equation (21) is a coupled first order differential equation that relates one spinor component to the derivative of the other. Solving it for $\chi_{\uparrow\downarrow}(y(x))$ gives the total spinor wavefunction as

$$\Psi(t,x)=a(x)^{-1}e^{i\varepsilon(q(x)-ct)}\chi(y(x)). \qquad (22)$$

Now, beside the solution of Eq. (21), the two functions $y(x)$ and $q(x)$ are essential ingredients in writing the total solution of the model in 1+1 space-time with static metric. Below, we refer to these as the "*characteristic functions*". Due to the fact that the volume element in curved space is proportional to the determinant of the metric, then normalization of the wavefunction requires that $\left|\sqrt{-g}\,\Psi(t,x)\right|$ be square integrable (with respect to the measure $dx$) where $g$ is the determinant of the metric. Since this determinant is $-(1+\alpha\beta)a(x)^2$, then the factor $a(x)^{-1}$ in the wavefunction assures normalization if $\left|\chi(y(x))\right|$ is square integrable.

To simplify the solution of Eq. (21) without diminishing the physical features of the problem, we take $\alpha=0$, which transforms the equation into

$$\left[\frac{d}{dy}+\begin{pmatrix}m & -E/\tau\\ E\tau & -m\end{pmatrix}\right]\begin{pmatrix}\chi_\uparrow(y)\\ \chi_\downarrow(y)\end{pmatrix}=0. \qquad (23)$$

Solving this equation for the two coupled components of the spinor wavefunction, $\chi_{\uparrow\downarrow}$, shows that both satisfy the following simple second order differential equation

$$\left[\frac{d^2}{dy^2}+\varepsilon^2-\left(m^2+\eta^2\right)+2i\varepsilon\eta\right]\chi_{\uparrow\downarrow}(y)=0. \qquad (24)$$

Moreover, the two components are relates as follows

$$\chi_\downarrow(y)=(\tau/E)\left(\tfrac{d}{dy}+m\right)\chi_\uparrow(y). \qquad (25a)$$

$$\chi_\uparrow(y)=(1/\tau E)\left(-\tfrac{d}{dy}+m\right)\chi_\downarrow(y). \qquad (25b)$$

We should now make the following four remarks about the solution. First, the effective mass of the spinor particle increases from $m$ in flat space to $M=\sqrt{m^2+\eta^2}$. Thus, we expect that $\eta^2$ be proportional to the curvature constant of the space. Second, the two boundaries of the curved space is obtained as solutions of the condition $g^{00}(x)=1-(1+\alpha\beta)b(x)^2=0$, which is usually satisfied by two points, say $X_\pm$, that constitute the boundaries of the space (i.e., $X_-\leq x\leq X_+$). Accordingly, one should require that the solution satisfy the boundary conditions $\chi(y(X_\pm))=0$. Now, in the flat space limit, we should expect that $X_\pm\to\pm\infty$ or that one of them is infinite and the other is zero (or finite that could always be shifted to the origin). Third, the solution is the product of a (growing or decaying) exponential $e^{\pm\omega y}$ and an oscillation $e^{\pm i\eta\varepsilon/\omega y}$, where $\omega$ is the amplitude "*decay parameter*" which is a positive function of the energy and reads as follows

$$\omega(\varepsilon) = \tfrac{1}{\sqrt{2}}\left[ m^2 + \eta^2 - \varepsilon^2 + \sqrt{\left(m^2 + \eta^2 - \varepsilon^2\right)^2 + (2\eta\varepsilon)^2} \right]^{1/2}. \tag{26}$$

For confined systems, the solution $\chi(y)$ is a linear combination of both $e^{\pm(\omega - i\eta\varepsilon/\omega)y}$ subject to the proper boundary conditions. However, for unconfined systems, the solution becomes the product of the oscillatory factor $e^{\pm i(\eta\varepsilon/\omega)y}$ and only a decaying factor $e^{-\omega|y|}$. Thus, $|\chi(y)|$ is square integrable assuring normalization of the total wave function $\Psi(t,x)$. Now, expression (26) shows that as the relativistic energy varies from $-\infty$ to $+\infty$, $\omega$ is bounded as $|\eta| \leq \omega(\varepsilon) \leq M$, where the minimum occurs as $|\varepsilon| \to \infty$ and the maximum at $\varepsilon = 0$. In part (a) of Figure 1, we give a graphical representation of this behavior where we plot $\omega$ as function of the energy for several values of the curvature parameter $\eta$. We also plot in part (b) the spatial oscillation "*wave number*", $|\eta|\varepsilon/\omega(\varepsilon)$. It is interesting to observe from Figure 1(a) that the distinction between bound ($\varepsilon^2 < m^2$) and unbound ($\varepsilon^2 > m^2$) states becomes insignificant for a strongly curved space: the wavefunctions for both states decay strongly (with almost the same rate) as the particle moves away from the origin. This observation is also confirmed in Figure 1(b) where the spatial oscillation wave number for large values of $\eta$ becomes linear with the energy independently of whether $\varepsilon^2$ is larger or smaller than $m^2$. The fourth and final remark is that the solution space splits into two disconnected subspaces. We call one of them the positive energy subspace and the other the negative energy subspace. The former (latter) is obtained by solving Eq. (24) for the top (bottom) component then substituting that in Eq. (25) to get the corresponding bottom (top) component. Doing so and substituting in Eq. (22), we obtain the positive energy subspace as linear combination of the following two spinor wavefunctions for $\pm y > 0$:

$$\Psi_\pm(t,x) = \frac{e^{\mp\omega y}}{a(x)} e^{i\varepsilon(q(x)-ct)} \begin{pmatrix} A_\pm e^{+i(\varepsilon/\omega)\eta y} + B_\pm e^{-i(\varepsilon/\omega)\eta y} \\ \frac{m\mp\omega+i\varepsilon\eta/\omega}{(\varepsilon+i\eta)/\tau} A_\pm e^{+i(\varepsilon/\omega)\eta y} + \frac{m\mp\omega-i\varepsilon\eta/\omega}{(\varepsilon+i\eta)/\tau} B_\pm e^{-i(\varepsilon/\omega)\eta y} \end{pmatrix}, \tag{27a}$$

where $A_\pm$ and $B_\pm$ are four independent normalization constants. On the other hand, the negative energy solution subspace is a linear combination of the following:

$$\Psi_\pm(t,x) = \frac{e^{\mp\omega y}}{a(x)} e^{i\varepsilon(q(x)-ct)} \begin{pmatrix} \frac{m\pm\omega-i\varepsilon\eta/\omega}{\tau(\varepsilon+i\eta)} A_\pm e^{+i(\varepsilon/\omega)\eta y} + \frac{m\pm\omega+i\varepsilon\eta/\omega}{\tau(\varepsilon+i\eta)} B_\pm e^{-i(\varepsilon/\omega)\eta y} \\ A_\pm e^{+i(\varepsilon/\omega)\eta y} + B_\pm e^{-i(\varepsilon/\omega)\eta y} \end{pmatrix}, \tag{27b}$$

where the signs also correspond to $\pm y > 0$. Since the Dirac equation is first order in the derivatives, then continuity is satisfied just by matching the solutions at the origin. That is, we require $\Psi_+(t,0^+) = \Psi_-(t,0^-)$, which results in two equations relating $(A_-, B_-)$ to $(A_+, B_+)$. Next, we use the above findings to write the solution of two illustrative problems.

### III. EXAMPLES IN THE FREE CASE

In this section, we obtain the two-component wavefunction for a spinor particle in 1+1 space-time with static metric influenced only by the gravitational background without coupling to any external potential. We present two examples of curved spaces where each one corresponds to a given function *f(x)* that leads to *a(x)* and *b(x)* as solutions of Eq. (17). The constraints stated in point (ii) below Eq. (17) lead to specific realizations of these two functions by restricting and/or fixing their associated parameters. Using $a(x)$ and $b(x)$ of the given problem, we compute the characteristic functions $y(x)$ and $q(x)$. Next, we solve Eq. (21) for the upper and lower spinor components $\chi_{\uparrow\downarrow}(y(x))$ in the positive and negative energy subspaces. The total wavefunction is then written as shown in Eq. (22). On the other hand, for $\alpha = 0$ the explicit representation of the wave function is already given by Eq. (27) where we only need to substitute the specific functions $a(x)$, $q(x)$ and $y(x)$.

### III.1 The case $f(x) = 0$:

With $f(x) = 0$, the solutions of Eq. (17) are $a(x) = (a_0 + a_1 x)/\sqrt{1+\alpha\beta}$ and $b(x) = (b_0 + b_1 x)/\sqrt{1+\alpha\beta}$. The flat space limit given by (8) dictates that $b_0 = 0$, $b_1 \to 0$ and $h(x) = a_0 + a_1 x$. Therefore, if $R$ is the curvature constant of the space then we can take $b_1 = \xi\sqrt{R}$, where $\xi$ is a dimensionless positive parameter. Substituting in $\Sigma_0 = \eta = (1+\alpha\beta)(ab' - a'b)$, we obtain $a_0 = \eta/\xi\sqrt{R}$ and to make it finite in the flat space limit we should require that $\eta$ be proportional to $\sqrt{R}$ (so that we also preserve the inverse length dimensionality of $\eta$). That is, we write $\eta = \theta\sqrt{R}$, where $\theta$ is a dimensionless *non-zero* positive parameter and thus $a_0 = \theta/\xi$. The constraint (16) gives $a_1 = \pm\theta\sqrt{R}$. Accordingly, the flat space metric is $g = \begin{pmatrix} 1 & 0 \\ 0 & -(\theta/\xi)^2 \end{pmatrix}$ and we end up with the following elements of the model:

$$a(x) = \theta\xi^{-1}(1\pm\xi\kappa x)/\sqrt{1+\alpha\beta}, \quad b(x) = \xi\kappa x/\sqrt{1+\alpha\beta}. \tag{28a}$$

$$g^{00} = 1 - (\xi\kappa x)^2, \quad g^{11} = -\theta^2\xi^{-2}(1\pm\xi\kappa x)^2, \quad g^{01} = g^{10} = -\eta x(1\pm\xi\kappa x). \tag{28b}$$

$$\Omega = \frac{\mp i\eta}{\sqrt{1+\alpha\beta}}\begin{pmatrix} 1 & \alpha \\ \beta & -1 \end{pmatrix} + \eta\begin{pmatrix} 0 & 1/\tau \\ \tau & 0 \end{pmatrix}, \tag{28c}$$

where $\kappa = \sqrt{R}$. Using these $a(x)$ and $b(x)$, we obtain the two characteristic functions $y(x)$ and $q(x)$ defined in the previous section as $\frac{1}{\sqrt{1+\alpha\beta}}\int\frac{dx}{a}$ and $\int\frac{b}{a}dx$, respectively. Performing these elementary integrations, we obtain $y_\mp(x) = \pm\eta^{-1}\ln(1\pm\xi\kappa x)$ and $q_\mp(x) = \frac{1}{\eta}[\pm\xi\kappa x - \ln(1\pm\xi\kappa x)] = \pm\left[\frac{\xi}{\theta}x - y_\mp(x)\right]$. As stated above, configuration space is bounded by the condition $g^{00}(x) = 0$, which translates into $-X < x < +X$, where $X = 1/\xi\sqrt{R}$. To make the argument of the log positive for all $x$, we associate the top (bottom) signs in the formulas above with $x < 0$ ($x > 0$) (i.e., with $\pm x < 0$). Thus, $x \in [-X, 0]$ corresponds to $y_-(x) \in [-\infty, 0]$ and $x \in [0, +X]$ corresponds to $y_+(x) \in [0, +\infty]$. With $y_\pm(x)$ and $q_\pm(x)$ being specified, the spinor wavefunction (22) becomes $\Psi_\pm(t,x) = \frac{e^{i\varepsilon(q_\pm(x)-ct)}}{a(x)\sqrt{1+\alpha\beta}}\chi(y_\pm)$, where $\chi(y_\pm)$ is the solution of Eq. (21). However, to simplify the solution without diminishing the general features of the problem, we take $\alpha = 0$. Substituting from the above results into Eq. (27a), we obtain the positive energy wavefunction for $\pm x > 0$ as linear combination of the following:

$$\Psi_{\pm}(t,x) = \frac{e^{\mp \omega y_{\pm}} e^{-i\varepsilon(ct\pm x\xi/\theta)} e^{\pm i\varepsilon y_{\pm}}}{\theta \xi^{-1}(1\mp \xi\kappa x)} \begin{pmatrix} A_{\pm} e^{+i(\varepsilon/\omega)\eta y_{\pm}} + B_{\pm} e^{-i(\varepsilon/\omega)\eta y_{\pm}} \\ \frac{m\mp\omega+i\varepsilon\eta/\omega}{(\varepsilon+i\eta)/\tau} A_{\pm} e^{+i(\varepsilon/\omega)\eta y_{\pm}} + \frac{m\mp\omega-i\varepsilon\eta/\omega}{(\varepsilon+i\eta)/\tau} B_{\pm} e^{-i(\varepsilon/\omega)\eta y_{\pm}} \end{pmatrix}, \quad (29a)$$

On the other hand, the negative energy solution subspace is obtained by substitution in Eq. (27b) giving the associated spinor wavefunction for $\pm x > 0$ as a linear combination of the following:

$$\Psi_{\pm}(t,x) = \frac{e^{\mp \omega y_{\pm}} e^{-i\varepsilon(ct\pm x\xi/\theta)} e^{\pm i\varepsilon y_{\pm}}}{\theta \xi^{-1}(1\mp \xi\kappa x)} \begin{pmatrix} \frac{m\pm\omega-i\varepsilon\eta/\omega}{\tau(\varepsilon+i\eta)} A_{\pm} e^{+i(\varepsilon/\omega)\eta y_{\pm}} + \frac{m\pm\omega+i\varepsilon\eta/\omega}{\tau(\varepsilon+i\eta)} B_{\pm} e^{-i(\varepsilon/\omega)\eta y_{\pm}} \\ A_{\pm} e^{+i(\varepsilon/\omega)\eta y_{\pm}} + B_{\pm} e^{-i(\varepsilon/\omega)\eta y_{\pm}} \end{pmatrix}. \quad (29b)$$

Matching the solutions at the origin gives $A_{-} = A_{+} + i\frac{\omega^2}{\varepsilon\eta}(A_{+}+B_{+})$ and $B_{-} = B_{+} - i\frac{\omega^2}{\varepsilon\eta}(A_{+}+B_{+})$. Aside from a constant factor and an oscillation factor of unit amplitude, the norm of this wavefunction at any point $x$ and energy $\varepsilon$ is bounded by $N(x,\varepsilon) = e^{\mp \omega y_{\pm}(x)}/(1\mp \xi\kappa x)$. As stated below Eq. (26), the minimum value of $\omega$ is $\eta$ and occurs at $\varepsilon = \pm\infty$ whereas the maximum value is $M$ that occurs at $\varepsilon = 0$. Inserting these values and $y_{\pm}(x)$ in the expression for $N(x,\varepsilon)$, we obtain $(1\mp \xi\kappa x)^{-1+\sqrt{1+m^2/\theta^2 R}} \leq N(x,\varepsilon) \leq 1$ for $x \in [0,\pm X]$. Aside from the rest mass $m$ and curvature constant $R$, this problem has three parameters: $\theta$, $\xi$ and $\tau$ (remembering that $\kappa = \sqrt{R}$ and $\eta = \theta\kappa$). The probability current vector is defined as $J^{\mu} = \bar{\Psi}\gamma^{\mu}\Psi$, where $\bar{\Psi} = \Psi^{\dagger}\gamma^{0}$. Therefore, the probability density is $\rho(x) = J^{0} = g^{00}(\Psi^{\dagger}\Psi)$. Using Eq. (29a), we plot in Figure 2 the positive energy probability density for a given set of physical parameters and for several energies.

### III.2 The case $f(x)$ = positive constant:

We write $f(x) = \zeta R$, where $\zeta$ is a positive dimensionless constant. In this case, the solution of Eq. (17) gives $\sqrt{1+\alpha\beta}\, a(x) = a_{+}\cosh(kx) + a_{-}\sinh(kx)$ and $\sqrt{1+\alpha\beta}\, b(x) = b_{+}\cosh(kx) + b_{-}\sinh(kx)$, where $k = \sqrt{\zeta R} = \kappa\sqrt{\zeta}$. The flat space limit ($k \to 0$) dictates that $b_{+} = -b_{-} \equiv b_{0}$ and gives $h(x) = a_{+}$. The equation $\Sigma_{0} = \eta$ gives $b_{0} = \eta/ka_{+}$ and the constraint (16) gives $a_{-}^2 = (1+b_{0}^2)a_{+}^2$. Putting all of this together and rewriting $a_{+}$ as $a_{0}$ and $\eta$ as $\vartheta k$, where $\vartheta$ is a dimensionless positive parameter, we obtain

$$\sqrt{1+\alpha\beta}\, a(x) = a_{0}\left[\cosh(kx) \pm \sqrt{1+(\vartheta/a_{0})^2}\,\sinh(kx)\right],$$
$$\sqrt{1+\alpha\beta}\, b(x) = (\vartheta/a_{0})\sinh(kx). \quad (30)$$

The boundaries of configuration space, which is obtained as solutions of the equation $g^{00}(x) = 0$, gives $X = \frac{1}{k}\left|\sinh^{-1}(a_{0}/\vartheta)\right|$ where $-X < x < +X$. If we define $\rho = \sqrt{1+(\vartheta/a_{0})^2}$ and $\sigma = \sqrt{\frac{\rho+1}{\rho-1}}$ then using the identity $\sinh^{-1}(x) = \ln\left(x+\sqrt{1+x^2}\right)$ gives $e^{kX} = \sigma$ and thus, $\sigma > e^{kx} > 1/\sigma$. For writing the two characteristic functions $y(x)$ and $q(x)$, we use the integral formulas [6]

$$\int \frac{e^x + \alpha e^{-x}}{e^x + \beta e^{-x}}\, dx = \frac{\alpha}{\beta} x + \frac{\beta-\alpha}{2\beta}\ln\left(\left|e^{2x}+\beta\right|\right), \quad (31a)$$

$$\int \frac{dx}{x^2-\alpha^2} = -\frac{1}{\alpha}\begin{cases} \tanh^{-1}(x/\alpha) &, x^2 < \alpha^2 \\ \tanh^{-1}(\alpha/x) &, x^2 > \alpha^2 \end{cases}, \tag{31b}$$

and with $a_0 > 0$ we associate the top (bottom) signs in formula (30) for $a(x)$ with $x < 0$ ($x > 0$) (i.e., with $\pm x < 0$) [7]. Consequently, we obtain $y_\mp(x) = \mp \frac{2a_0}{\eta\vartheta}\big[\tanh^{-1}(e^{\mp kx}/\sigma) -$ $\tanh^{-1}(1/\sigma)\big]$ and $q_\mp(x) = \frac{\vartheta/2ka_0^2}{\pm\rho-1}\big[2kx + (\sigma^{\mp 2}-1)\ln(\pm e^{2kx} \mp \sigma^{\mp 2})\big]$ where both correspond to $\pm x < 0$. Note that in the definition of $y(x)$ we chose a constant of integration just to make $y_\pm(0) = 0$. With these specific realizations of the characteristic functions, the spinor wavefunction (22) becomes $\Psi_\pm(t,x) = \frac{e^{i\varepsilon(q_\pm(x)-ct)}}{a(x)\sqrt{1+\alpha\beta}}\chi(y_\pm(x))$, where $\chi(y_\pm)$ is the solution of Eq. (21) with $E = \varepsilon + i\vartheta k$. If we simplify by taking $\alpha = 0$ then we end up with the solution given by Eq. (27) with $(y,q) \to (y_\pm, q_\pm)$ for $\pm x > 0$. Figure 3 is a plot of the positive energy probability density for a given set of physical parameters and several energies. Note that in the limit $\zeta \to 0$, the solution in this subsection becomes equal to that in the previous subsection where $f(x) = 0$. Moreover, if $f(x)$ were equal to a negative rather than positive constant, then the functions $a(x)$ and $b(x)$ will be written in terms of trigonometric rather than hyperbolic functions.

We end this section by making the following two remarks about possible ways of obtaining the model functions $a(x)$ and $b(x)$. First, it is interesting to note that Eq. (17) is identical to the Schrödinger equation in ordinary one-dimensional quantum mechanics for a potential function $V(x) = \frac{1}{2}f(x)$ at zero energy. Therefore, all known solutions of such problems could be used in writing $a(x)$ and $b(x)$. For example, we may consider the inverse square potential case where $V(x) = -\frac{\mu}{2}x^{-2}$. Despite the fact that this potential is known to be plagued with quantum anomalies [8], it has a special interesting solution at zero energy when the dimensionless coupling parameter $\mu$ assumes the critical value $\frac{1}{4}$. The two independent solutions in this case are $\sqrt{x}$ and $\sqrt{x}\ln(x)$. Therefore, we can write $a(x) = \sqrt{\kappa x}[a_0 + a_1 \ln(\kappa x)]$ and $b(x) = \sqrt{\kappa x}[b_0 + b_1 \ln(\kappa x)]$, where $\kappa = \sqrt{R}$ and $x \geq 0$. Second, using $a'' = a'\frac{da'}{da}$ transforms Eq. (17) into $a'\frac{da'}{da} = fa$ whose integration yields $a'^2 = 2\int af(a)da$. If we define this integral as $Rg(a)^2$, then we obtain $x + x_0 = \pm\int da/\kappa g(a)$, where $x_0$ is an arbitrary constant. Assuming that these integrals are doable, then this constitutes a double integration technique for solving the second order differential equation (17). For example, taking $f(a) = \text{constant} = R$ gives $g(a) = a$ resulting in $a(x)$ being a linear combinations of $e^{\pm\kappa x}$, which is the case discussed in subsections III.2. Another example is for $f(a) \propto a^n$, where $n \neq 0$ giving $a(x) \propto x^{-2/n}$, which for $n = -4$ corresponds to the case of the inverse square potential at zero energy but without the log term.

–9–

## IV. THE MODEL WITH INTERACTION

We couple the Dirac spinor to the most general time-independent potentials in 1+1 space-time. These include scalar, pseudo-scalar, and vector potential couplings, which could be accomplished by the following replacements in the free Dirac equation (3):

$$\partial_\mu \to \partial_\mu + iA_\mu, \quad m \to m + S + iW\gamma^5, \tag{32}$$

where $\gamma^5 = i\gamma^0\gamma^1$, $S$ is the scalar potential, $W$ is the pseudo-scalar potential and $A_\mu$ is the two-vector potential. Writing the spinor wavefunction as $\Psi(t,x) = a(x)^{-1} e^{i\varepsilon(q(x)-ct)}\phi(x)$ and inserting the gamma matrices from Eq. (9) and the matrix $\Omega$ from Eq. (15) into the Dirac equation (3) with the replacement (32) results in the following matrix equation

$$\left\{ a\left[\frac{d}{dx} + i\left(A_1 + \frac{b}{a}A_0\right)\right] + (m+S+abW)\begin{pmatrix} 1 & 0 \\ 0 & -1 \end{pmatrix} + (A_0 - E)\begin{pmatrix} 0 & 1/\tau \\ -\tau & 0 \end{pmatrix} + iaW\begin{pmatrix} 0 & 1/\tau \\ \tau & 0 \end{pmatrix} \right\}\begin{pmatrix} \phi_\uparrow \\ \phi_\downarrow \end{pmatrix} = 0, \tag{33}$$

where $E = \varepsilon + i\eta$, $q(x)$ is defined as above, $\frac{dq}{dx} = b/a$, and we have chosen $\alpha = 0$. Writing $\phi(x) = e^{-ip(x)}\chi(x)$, where $\frac{dp}{dx} = A_1 + \frac{b}{a}A_0$, transforms this equation into the following one whose free version is Eq. (23)

$$\left[\frac{d}{dy} + \begin{pmatrix} m+S+abW & \tau^{-1}(A_0 - E + iaW) \\ \tau(E - A_0 + iaW) & -(m+S+abW) \end{pmatrix}\right]\begin{pmatrix} \chi_\uparrow(y) \\ \chi_\downarrow(y) \end{pmatrix} = 0, \tag{34}$$

where $y(x)$ is defined as above $\frac{dy}{dx} = 1/a$. One should note that this equation does not depend on the space component of the vector potential $A_1$, which is expected since it can always be gauged away by a $U(1)$ gauge transformation (here, it is $e^{-ip(x)}$). Defining $\mathcal{A} = m + S + abW$, $\mathcal{B} = E - A_0$ and $\mathcal{C} = aW$ we obtain the following set of equations for the positive energy subspace

$$\left[\frac{d^2}{dy^2} - \left(\frac{\dot{\mathcal{B}} - i\dot{\mathcal{C}}}{\mathcal{B} - i\mathcal{C}}\right)\left(\frac{d}{dy} + \mathcal{A}\right) + \dot{\mathcal{A}} - \mathcal{A}^2 + \mathcal{B}^2 + \mathcal{C}^2\right]\chi_\uparrow = 0, \tag{35a}$$

$$\chi_\downarrow = \frac{\tau}{\mathcal{B} - i\mathcal{C}}\left(\frac{d}{dy} + \mathcal{A}\right)\chi_\uparrow, \tag{35b}$$

where the dot over a variable stands for its derivative with respect to $y$. On the other hand, the negative energy states satisfy the following equations

$$\left[\frac{d^2}{dy^2} - \left(\frac{\dot{\mathcal{B}} + i\dot{\mathcal{C}}}{\mathcal{B} + i\mathcal{C}}\right)\left(\frac{d}{dy} - \mathcal{A}\right) - \dot{\mathcal{A}} - \mathcal{A}^2 + \mathcal{B}^2 + \mathcal{C}^2\right]\chi_\downarrow = 0, \tag{36a}$$

$$\chi_\uparrow = \frac{\tau^{-1}}{\mathcal{B} + i\mathcal{C}}\left(-\frac{d}{dy} + \mathcal{A}\right)\chi_\downarrow. \tag{36b}$$

As an example, we consider the problem with the potential configuration: $A_0(x) = 0$, $W(x) = W_0/a(x)$ and $S(x) = S_0 b(x)$, where $S_0$ and $W_0$ are real potential parameters. As a result, we obtain

$$\left[\frac{d^2}{dy^2} \pm (S_0 + W_0)\dot{b} - [m + (S_0 + W_0)b]^2 + W_0^2 + \varepsilon^2 - \eta^2 + 2i\varepsilon\eta\right]\chi_{\uparrow\downarrow}(y) = 0, \tag{37}$$

where the top (bottom) sign corresponds to the left (right) arrow. For the curved space studied in subsection III.1 where $a(x) = \theta \xi^{-1}(1 \mp \xi \kappa x)$ and $b(x) = \xi \kappa x$, the scalar potential becomes linear and the pseudo-scalar shifted Coulomb-like. Moreover, $\dot{b} = \xi \kappa \frac{dx}{dy} = \xi \kappa a$ and since $y_\pm(x) = \mp \eta^{-1} \ln(1 \mp \xi \kappa x)$, we obtain $\xi \kappa x = \pm(1 - e^{\mp \eta y_\pm})$ for $\pm x > 0$. Therefore, $\dot{b} = \eta e^{\mp \eta y_\pm}$ and Eq. (37) becomes as follows for $\pm x > 0$

$$\left[ \frac{d^2}{dy_\pm^2} - V_0^2 e^{\mp 2\eta y_\pm} + V_0 \left[ 2(V_0 \pm m) + \eta \right] e^{\mp \eta y_\pm} + \varepsilon^2 - \mathcal{M}_\pm^2 + 2i\varepsilon\eta \right] \chi_\uparrow(y) = 0, \tag{38a}$$

$$\left[ \frac{d^2}{dy_\pm^2} - V_0^2 e^{\mp 2\eta y_\pm} + V_0 \left[ 2(V_0 \pm m) - \eta \right] e^{\mp \eta y_\pm} + \varepsilon^2 - \mathcal{M}_\pm^2 + 2i\varepsilon\eta \right] \chi_\downarrow(y) = 0, \tag{38b}$$

where $V_0 = S_0 + W_0$ and $\mathcal{M}_\pm^2 = (m \pm V_0)^2 + \eta^2 - W_0^2$. This differential equation is that of the one-dimensional *generalized* Morse oscillator, $V(x) = V_2 e^{-2\mu x} - V_1 e^{-\mu x}$, which is solvable in terms of the confluent hypergeometric function. Defining $z = e^{\mp \eta y_\pm}$ transforms Eq. (38a) into the following one in terms of the variable $z$

$$\left( z \frac{d^2}{dz^2} + \frac{d}{dz} - \upsilon_2^2 z - \frac{\upsilon_0}{z} + \upsilon_1 \right) \chi_\uparrow(z) = 0, \tag{39}$$

where $\upsilon_2 = V_0/\eta$, $\upsilon_1 = \upsilon_2 \left[ 2(\upsilon_2 \pm m/\eta) + 1 \right]$ and $\upsilon_0 = (\mathcal{M}_\pm/\eta)^2 - (\varepsilon/\eta)^2 - 2i(\varepsilon/\eta)$. For Eq. (38b) we obtain the same equation as (39) but with $\upsilon_1 = \upsilon_2 \left[ 2(\upsilon_2 \pm m/\eta) - 1 \right]$. Now, if we write $\chi_\uparrow(z) = z^\nu e^{-z/2} H(z)$ where $\nu$ is a constant parameter, then Eq. (39) becomes the following equation for $H(z)$

$$\left[ z \frac{d^2}{dz^2} + (2\nu + 1 - z) \frac{d}{dz} + \frac{\nu^2 - \upsilon_0}{z} + \left( \tfrac{1}{4} - \upsilon_2^2 \right) z + \upsilon_1 - \left( \nu + \tfrac{1}{2} \right) \right] H(z) = 0. \tag{40}$$

Comparing this with the differential equation of the confluent hypergeometric function, $\left[ z \frac{d^2}{dz^2} + (\beta - z) \frac{d}{dz} - \alpha \right] {}_1F_1(\alpha; \beta; z) = 0$, we conclude that $H(z) = {}_1F_1\left(\nu + \tfrac{1}{2} - \upsilon_1; 2\nu + 1; z\right)$ provided that $\upsilon_2^2 = \tfrac{1}{4}$ and $\nu^2 = \upsilon_0$. Therefore, $V_0^2 = (\eta/2)^2$ which implies that in this system the curvature constant, which is proportional to $\eta^2$, is related to the sum of the potential strengths $S_0 + W_0$. Moreover, if we write $\nu = \pm \sqrt{\upsilon_0} = \pm(\omega - i\eta\varepsilon/\omega)$, then $\omega(\varepsilon)$ will be the same positive energy function as in Eq. (26) but with the replacement $m^2 \to (m \pm V_0)^2 - W_0^2$. With these results, we write the upper spinor component in the positive energy solution subspace as linear combination of the following two solutions with $\nu = \pm(\omega - i\eta\varepsilon/\omega)$

$$\chi_\uparrow(z) = z^\nu e^{-z/2} {}_1F_1\left( \nu + \tfrac{1}{2} - \upsilon_1; 2\nu + 1; z \right). \tag{41}$$

Proper boundary conditions will determine the right combinations. Now, since $z = e^{\mp \eta(\omega - i\eta\varepsilon/\omega) y_\pm}$ for $\pm x > 0$ and $\pm y_\pm > 0$ then $|z| \leq 1$ and the confluent hypergeometric series will converge for all values of $\nu$ (i.e., for all values of the energy). Therefore, $|\chi_\uparrow(z)|$ is square integrable and the total wavefunction is normalizable. The lower component for the positive energy subspace is obtained from the upper component (41) using the kinetic balance relation (35b), which reads in terms of the variable $z$ as follows

$$\chi_\downarrow(z) = \frac{\pm \tau}{\varepsilon + i(\eta - W_0)} \left[ \pm m + V_0 - z\left( \eta \frac{d}{dz} + V_0 \right) \right] \chi_\uparrow(z), \tag{42}$$

−11−

for $\pm x > 0$. In this calculation, we use the differential property and Recurrence relation of the confluent hypergeometric function [9]. With $\chi_{\uparrow\downarrow}(z)$ being finally specified, we write the total spinor wavefunction for the positive energy subspace as

$$\Psi_{\uparrow\downarrow}(t,x) = a(x)^{-1} e^{i\varepsilon(q(x)-ct)} e^{-ip(x)} \chi_{\uparrow\downarrow}(z(x)). \tag{43}$$

The lower component for the negative energy subspace is the same as the upper in (41) except that the parameter $\upsilon_1$ is defined with its last term equal to $-\upsilon_2$ not $+\upsilon_2$. The upper component is obtained from that by using relation (36b).

## V. CONCLUSION

We introduced a one-parameter model of the Dirac equation in a curved space by requiring compatibility of symmetric quantization of the Dirac wave operator with general covariance. We also imposed the condition that the square of the equation gives the conventional Klein-Gordon equation in curved space. The result is a special representation where the dot product of the spin connections with the Dirac gamma matrices is equal to the "covariant divergence" of the latter. Additionally, the gamma matrices are required to satisfy a matrix operator algebra involving the connections in the "covariant derivative" and the Riemann-Christoffel connections. The model was worked out explicitly in Sec. II for the case of a curved space with static metric and in Sec. III we solved it for few illustrative examples. We also formulated the interacting model and solved it exactly for a given physical configuration. We believe that the matrix operator algebra introduced here constitutes stringent requirements that pin down the representation of the Dirac gamma matrices and avoid the non-uniqueness problem in formulating the Dirac equation in a curved space.

## ACKNOWLEDGEMENT

We appreciate the generous support provided by the Saudi Center for Theoretical Physics (SCTP). We also acknowledge partial support by King Fahd University of Petroleum & Minerals (KFUPM).

**REFERENCES:**

[1] A. D. Alhaidari and A. Jellal, "*Compatibility of symmetric quantization with general covariance in the Dirac equation and spin connections*", Phys. Lett. A **379** (2015) doi:10.1016/j.physleta.2015.08.021, in production.

[2] H. Weyl, "*Elektron und Gravitation*", Z. Phys. **56**, 330 (1929); V. A. Fock, "*Geometrisierung der Diracschen Theorie des Elektrons*", Z. Phys. **57**, 261 (1929).

[3] H. S. Green, "*Dirac matrices, teleparallelism and parity conservation*", Nucl. Phys. **7**, 373 (1958).

[4] H. A. Weldon, "*Fermions without vierbeins in curved space-time*", Phys. Rev. D **63**, 104010 (2001).

[5] M. Arminjon and F. Reifler, "*A non-uniqueness problem of the Dirac theory in a curved spacetime*", Ann. Phys. (Berlin) **523**, 531 (2011); M. Arminjon, "*On the Non-uniqueness Problem of the Covariant Dirac Theory and the Spin-Rotation

*Coupling*", Int. J. Theor. Phys. **52**, 4032 (2013); M. Arminjon, "*Some Remarks on Quantum Mechanics in a Curved Spacetime, Especially for a Dirac Particle*", Int. J. Theor. Phys. **54**, 2218 (2013) and references therein.

[6] We also used the following identities:
  (i) $\ln\left(\frac{x-\alpha}{x+\alpha}\right) = -2\coth^{-1}(x/\alpha)$ for $x^2 > \alpha^2$, $\ln\left(\frac{x-\alpha}{x+\alpha}\right) = -2\coth^{-1}(x/\alpha)$ for $x^2 > \alpha^2$.
  (ii) $\coth^{-1}(x/\alpha) = \tanh^{-1}(\alpha/x)$ for $x^2 > \alpha^2$, $\coth^{-1}(\alpha/x) = \tanh^{-1}(x/\alpha)$ for $x^2 < \alpha^2$.

[7] For negative $a_0$, we make the signs association with $\pm x > 0$ not with $\pm x < 0$ and in the definition of $y(x)$ and $q(x)$ we make the replacement $y_\mp \to y_\pm$ and $q_\mp \to q_\pm$.

[8] D. Bouaziz and M. Bawin, "*Singular inverse-square potential: Renormalization and self-adjoint extensions for medium to weak coupling*", Phys. Rev. A **89**, 022113 (2014) and references therein.

[9] These are: $\frac{d}{dz} {}_1F_1(\alpha;\beta;z) = \frac{\alpha}{\beta} {}_1F_1(\alpha+1;\beta+1;z)$, and

$z\,{}_1F_1(\alpha;\beta;z) = (\beta-2\alpha)\,{}_1F_1(\alpha;\beta;z) + (\alpha-\beta)\,{}_1F_1(\alpha-1;\beta;z) + \alpha\,{}_1F_1(\alpha+1;\beta;z)$,

$z\,{}_1F_1(\alpha;\beta;z) = (1-\alpha)\,{}_1F_1(\alpha;\beta;z) + (\alpha-\beta)\,{}_1F_1(\alpha-1;\beta;z) + (\beta-1)\,{}_1F_1(\alpha;\beta-1;z)$.

**FIGURE CAPTIONS:**

**Fig. 1**: (a) Plot of the amplitude "decay parameter" of spatial oscillation, $\omega$, as function of the relativistic energy $\varepsilon$ for several values of the curvature parameters $\eta$, all in units of the rest mass $m$. (b) Plot of the spatial oscillation "wave number" $|\eta|\varepsilon/\omega(\varepsilon)$. In both plots, traces with (empty squares, empty circles, filled squares, filled circles) correspond to the values (0.15, 0.4, 0.8, 1.5) for $\eta$. The thick solid trace with no attached symbols corresponds to the flat space limit where we took $\eta = 0.001$.

**Fig. 2**: The positive energy probability density associated with the solution in Eq. (29a) for the space with a curvature constant $R = 0.2m^2$ and physical parameters $\{\theta,\xi,\tau\} = \{0.3,0.5,1.0\}$ and we took $\{A_+,B_+\} = \{1.2,0.8\}$. The traces with (squares, circles, triangles) correspond to values of $\varepsilon$ equal to (0.8, 1.0, 1.2) in units of the rest mass. The thick solid trace is for $\varepsilon = 0.5m$. The x-axis was scaled by the boundary value $X$ and, for better presentation, the plots were scaled to bring their maxima to almost the same height.

**Fig. 3**: The positive energy probability density associated with the solution of the problem in subsection III.2 for $R = 0.2m^2$ and $\{\zeta,\vartheta,a_0,\tau\} = \{0.7,0.3,0.5,1.0\}$. The normalization constants and energy values are the same as in Fig. 2.

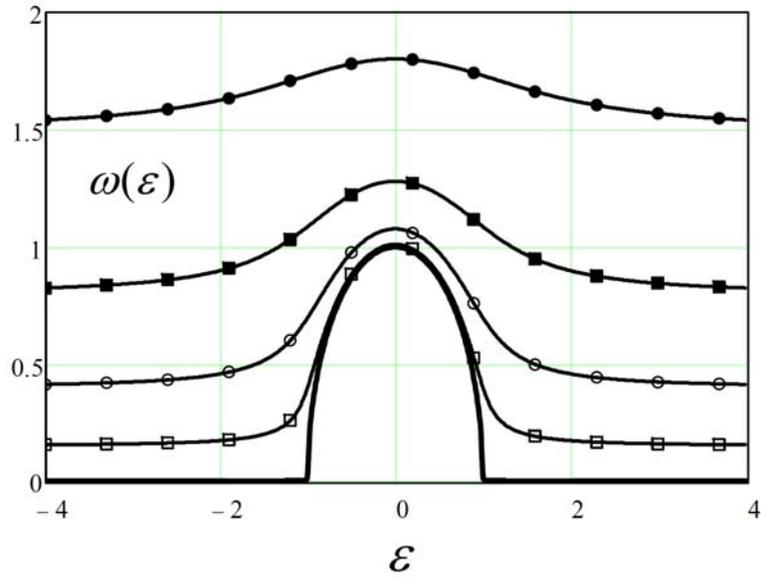

**Fig. 1(a)**

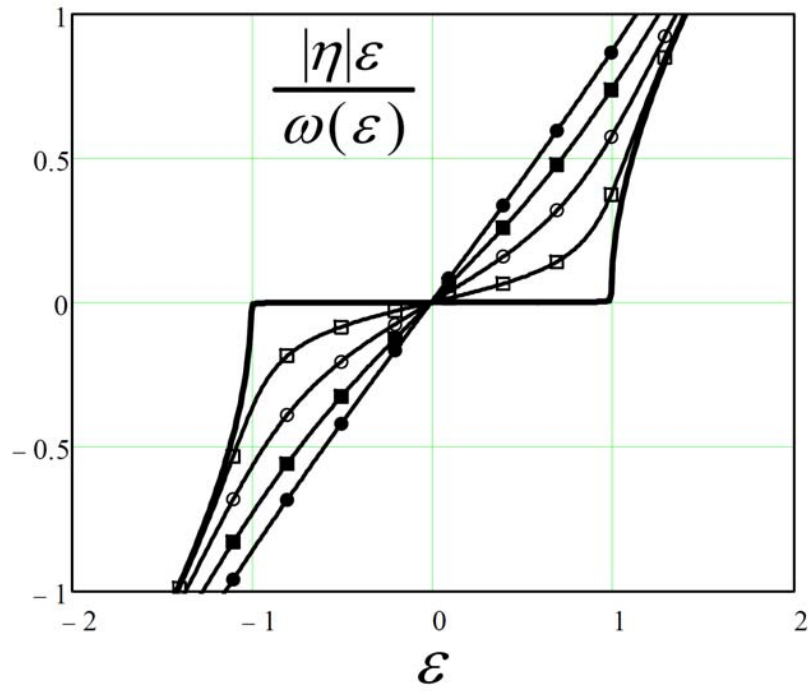

**Fig. 1(b)**

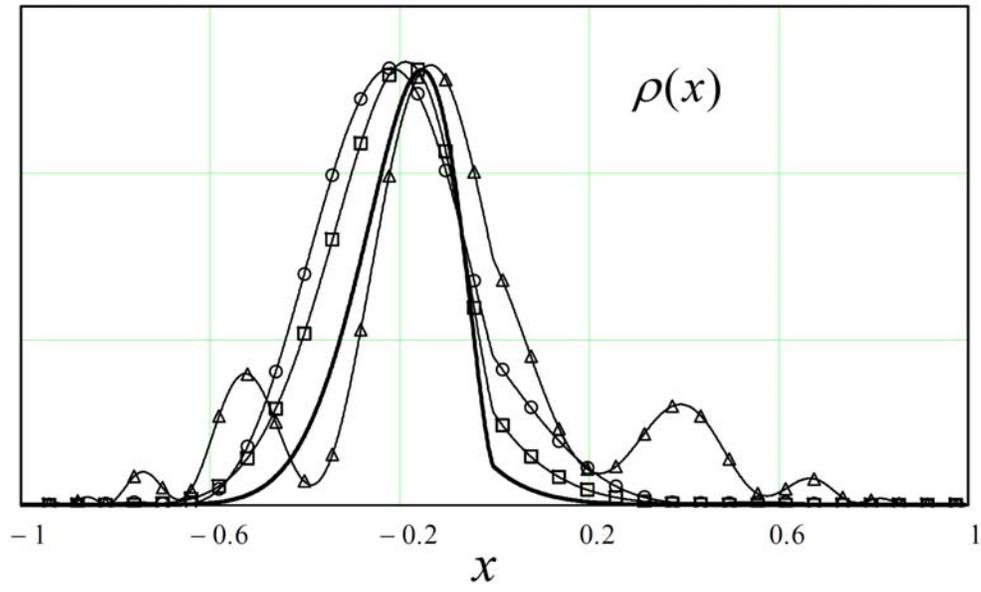

**Fig. 2**

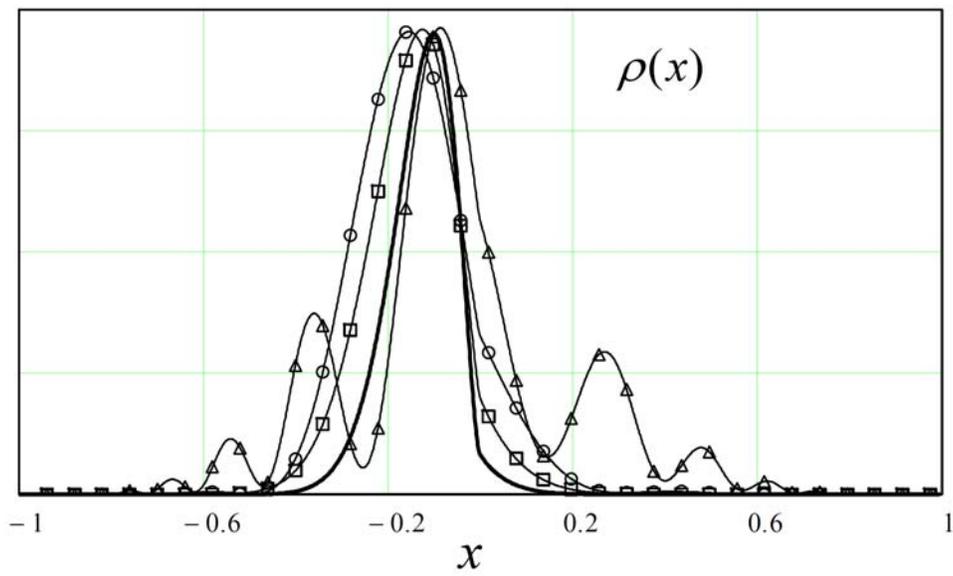

**Fig. 3**